\documentclass[11pt]{article}
\usepackage{epsfig}
\def\prl {Phys. Rev. Lett. }
\def\pr{Phys. Rev. }
\def\np{Nucl. Phys. }

\newcommand{\ed}{\end{document}}
\newcommand{\bq}{\begin{equation}}
\newcommand{\eq}{\end{equation}}
\newcommand{\ba}{\begin{eqnarray}}
\newcommand{\ea}{\end{eqnarray}}
\newcommand{\baz}{\begin{eqnarray*}}
\newcommand{\eaz}{\end{eqnarray*}}

\newcommand{\ben}{\begin{enumerate}}
\newcommand{\een}{\end{enumerate}}
\begin{document}

\hfill {FTUV-02-0103 IFIC-02-01}

\hfill {KIAS-P01071}

\hfill {SNUTP/02001}

\hfill {\today: hep-ph/}

\begin{center}
\ \\
{\Large \bf Atiyah-Manton Approach to Skyrmion Matter} \\

\vspace{1cm} {Byung-Yoon Park}

{\it Department of Physics, Chungnam National University,
Daejon 305-764,
Korea}\\ ({\small E-mail: bypark@chaosphys.chungnam.ac.kr})

\vskip 0.2cm
{Dong-Pil Min }

{\it Department of Physics and Center for Theoretical
Physics}\\
{\it Seoul National University, Seoul 151-742, Korea}\\
({\small
E-mail: dpmin@phya.snu.ac.kr})
\vskip 0.2cm
{Mannque Rho}

{\it Service de Physique Th\'eorique, CE Saclay}\\ {\it
91191
Gif-sur-Yvette, France}\\ {\&}\\{\it School of Physics,
Korea
Institute for Advanced Study, Seoul 130-012, Korea}\\
({\small
E-mail: rho@spht.saclay.cea.fr})

\vskip 0.2cm
{Vicente Vento }

{\it Departament de Fisica Te\`orica and Institut de
F\'{\i}sica
Corpuscular}\\ {\it Universitat de Val\`encia and Consejo
Superior
de Investigaciones Cient\'{\i}ficas}\\ {\it E-46100
Burjassot
(Val\`encia), Spain}\\ {\&}\\{\it School of Physics,
Korea
Institute for Advanced Study, Seoul 130-012, Korea}\\({\small E-mail:
vicente.vento@uv.es})

\end{center}
\vskip 0.5cm \centerline{\bf Abstract}

\vskip 0.4cm

We propose how to approach, and report on the first results in our
effort for, describing nuclear matter starting from the solitonic
picture of baryons which is supposed to represent QCD for large
number of colors. For this purpose, the instanton-skyrmion
connection of Atiyah and Manton is exploited to describe skyrmion
matter. We first modify 't Hooft's multi-instanton solution so as
to suitably incorporate proper dynamical variables into the
skyrmion matter and then by taking these variables as variational
parameters, we show that they cover a configuration space
sufficient to adequately describe the ground state properties of
nuclear matter starting from the skyrmion picture. Our results
turn out to be comparable to those so far found in different
numerical calculations, with our solution reaching stability at
high density for a crystal structure and obtaining a comparable
value for the energy per baryon at the minimum, thus setting the
stage for the next step.

\newpage
\section{Introduction}
Understanding from ``first principles" the ground state of
multi-baryon matter, namely, nuclear matter, is currently one of
the most important issues in nuclear physics in connection with
the mapping of the QCD phase structure. It is indispensable to the
physics of hadronic matter under extreme conditions such as what
is believed to be encountered in the interior of compact stars and
in relativistic heavy-ion collisions. Of various approaches
developed so far for the problem, one can cite, broadly, two
classes of approaches to the problem: One deals directly with QCD
variables and the other indirectly although both are anchored on
the modern theory of strong interactions, QCD.

In the first class is the approach that deals directly with the
quark and gluon fields but in a nonperturbative
setting~\cite{berges}. This is emerging in the most recent
development and being in a seminal stage, does not yet lend itself
to a direct confrontation with nature. We will have little to say
on this in this paper. Let it suffice to briefly state that it
exploits the possible ``quark-hadron continuity" observed at high
density~\cite{continuity} and extrapolated downward to low density
in the ``top-down" way.

In the second class are two approaches that resort to effective
field theories of QCD that deal with $effective$ degrees of
freedom, namely, hadronic fields. Of the two, the one closest in
spirit to the standard nuclear physics approach (SNPA) is based on
chiral dynamics given in terms of local baryonic (nucleon,
$\Delta$, etc) and mesonic (pion, vector meson $\rho$, $\omega$,
etc.) fields. Here nuclear matter emerges from an effective chiral
$action$ as a sort of Q-matter or ``chiral-liquid" nontopological
soliton~\cite{lynn}. This description can then be related to
Landau Fermi liquid structure~\cite{migdal} via Walecka's
model~\cite{walecka} when the notion of ``sliding vacuum" in dense
medium is implemented as discussed in \cite{br91,br01}. This
approach allows a contact with nature and suggests how to
extrapolate to densities higher than that of nuclear matter. Some
of the recent developments along this line are reviewed by
Song~\cite{song}. While this approach relies on an effective field
theory strategy, thereby maintaining a link to ``first principles"
and enabling certain phase transitions like kaon condensation to
be described within the same framework~\cite{kaon}, numerous
assumptions that are not directly accountable by the fundamental
theory are invoked to arrive at the forms used for describing
nature. As it stands, it is difficult to say to what extent this
approach represents QCD and how reliable it is in confronting
nature.

The other method in this second class which is in principle
closest to QCD is the topological soliton approach based on an
effective Lagrangian. In principle, given an effective chiral
Lagrangian that is as ``accurate" as can be, one should be able to
quantitatively describe $n$-baryon systems for arbitrary $n$
including nuclear matter as $solitons$ of the Lagrangian.
Presently we are far from being able to do this for the obvious
reasons: First of all, such a Lagrangian is not available and
secondly, even if it were available, we would not know how to
obtain reliable soliton solutions. In the absence of an indication
that this feat can be realized in the immediate future, our
present aim is to use a ``reasonable" but simple Lagrangian and
arrive at a QCD-based description of nuclear matter that can be
probed for any given density. The objective of this paper is to
make a first such step toward that goal which is to pick a simple
Lagrangian and develop a scheme which can be exploited to reach
the ultimate goal, i.e., nuclear matter via skyrmion.

What we accomplish in this paper is admittedly quite modest but we
feel that it presents a promising way to make the next step and
hence deserves to be discussed. For this purpose, we shall take
the simplest such effective Lagrangian known that is shown to be
semi-realistic, namely, the Skyrme model~\cite{skyrme}.

The Skyrme model describes the baryons with an arbitrary baryon
number as static soliton solutions of an effective Lagrangian for
pions. This model has been used to describe not only single baryon
properties\cite{ANW83,ZB86}, but also has served to derive the
nucleon-nucleon interaction\cite{skyrme,JJP85}, the pion-nucleon
interaction\cite{SWHH89}, properties of light nuclei and of
nuclear matter. In the case of nuclear matter, most of the
developments \cite{KPR84,Kl85,BJW85,GM87,JV88,CJJVJ89,KS89,MS95}
involve the skyrmion crystal,  except for  a few
exceptions\cite{Ka98} where the skyrmion fluid is considered. The
first trial to understand dense skyrmion matter was made by
Kutchera {\it et al.}\cite{KPR84}. In their work, a single
hedgehog skyrmion is put into a spherical Wigner-Seitz cell
without incorporating explicit information on the interaction.

The conventional approach so far studied to skyrmion matter is to
assume a crystal symmetry and then to perform numerical
simulations using the symmetry as a constraint. The first guess at
this symmetry was made by Klebanov \cite{Kl85}. He considered a
system where the skyrmions are located in the lattice site of a
cubic crystal (CC) and have relative orientations in such a way
that the pair of nearest neighbors attract maximally. Goldhaber
and Manton \cite{GM87} suggested that contrary to Klebanov's
findings, the high density phase of skyrmion matter is to be
described by a body-centered lattice (BCC) of half skyrmions. This
suggestion was shown to be confirmed \cite{JV88} and its energy
per baryon of the ground state calculated, $(E/B)_{min} = 1.076 \;
({6\pi^2 f_\pi}/{e})$\footnotemark when the lattice size($L$) is
$L_{\min}= 5.56 \;  ({\hbar c}/{e f_\pi})$. \footnotetext{We use
throughout the paper the units in parenthesis to express the
energy and the length. In these expressions $f_\pi$, the pion
decay constant, and $e$ are parameters appearing in  Skyrme's
model \cite{skyrme}, as we shall show in the next section. We
recall also that the energy of a single skyrmion is -- in these
units -- 1.23 in the conventional parameterization \cite{ANW83}.}

Kugler and Shtrikman \cite{KS89}, using a variational method,
investigated the ground state of the skyrmion crystal including
not only the single skyrmion CC and half-skyrmion BCC but also the
single skyrmion face-centered-cubic (FCC) and half-skyrmion CC.
The ground state was found to be a half-skyrmion CC with the
energy per baryon of $(E/B)_{min} = 1.038$ at $L_{\min}= 4.72$.
This simple cubic arrangement of half skyrmions undergoes a
further phase transition at very high densities to a BCC crystal
of half skyrmions.

These classical crystalline structures are quite far from normal
nuclear matter which is known to be in a Fermi liquid phase at low
temperature\cite{migdal,NM}.
In order for the skyrmion matter to be identified with nuclear
matter, we have to quantize and thermalize the classical system.
Since it is a system of solitons in a pion field theory, it is not
sufficient to quantize the pion fields only. We have to introduce
and quantize proper collective variables {\em not only} to
complement the broken symmetries of the whole skyrmion system {\em
but also} to describe the dynamics of the single skyrmions in
order to obtain a realistic picture of nuclear matter. The
limitation of the works that include only the former has been
discussed in Ref.\cite{Co89}. In order to describe a system of
extended objects, e.g. skyrmions, we need a large number of
dynamical variables, such as, the position of their center of
mass, their relative orientations, their size, their deformation,
etc. Among them, those degrees of freedom that describe
translations and rotations of the single skyrmion play the
dominant role at low energy. Thus we need at least 6 variables for
each skyrmion. For a single skyrmion, the collective variables
that define the orientation in isospin space are quantized to give
rise to nucleons and isobars\cite{ANW83}. For the multi-skyrmion
system, the simplest way of introducing collective variables for
the position and orientation of each single skyrmion is through
the use of the product ansatz, the old idea of Skyrme to
investigate the nucleon-nucleon force\cite{skyrme,JJP85}. In this
case, a multi-skyrmion solution can be obtained by products of
single skyrmion solutions centered at the corresponding positions
and rotated to have the corresponding orientation. However, the
product ansatz  works well only when the skyrmions are
sufficiently separated. Furthermore, due to the non-commutativity
of the matrix products, it is very difficult to apply the product
ansatz to many-skyrmion systems.\footnote{The symmetrization
principle haunts us: we have to construct a system of identical
particles, at the classical level, such that the physical
quantities are invariant under the exchange of any two of them.}

Another scheme which has been used to study multi-skyrmion systems
is the procedure based on the Atiyah-Manton ansatz \cite{AM89}. In
this scheme, skyrmions of baryon number $N$ are obtained by
calculating the holonomy of Yang-Mills instantons of charge $N$.
This ansatz has been used successfully in few-nucleon systems
\cite{AM93,LM94,LMS95,Wa94,Wa96}.  This procedure has been also
applied to nuclear matter with the instanton solution on a four
torus \cite{MS95}. The energy per baryon was found to be
$(E/B)_{min} = 1.058$ at $L_{\min}= 4.95$, which is comparable to
the variational result of Kugler and Shtrikman {\it et
al\/}\cite{KS89}. One advantage of the Atiyah-Manton ansatz over
others is that it provides a natural framework to introduce the
proper dynamical variables for the skyrmions through the
parameters determining the multi-instanton configuration. Several
useful ans\"{a}tze for multi-instanton solutions are available in
the literature\cite{tHooft,JNR77}. Contrary to the {\em product}
rule of the  multi-skyrmion solution as a product ansatz, the
multi-instanton solutions are given in terms of an {\em additive}
rule.

In our first effort to go from a skyrmion matter to nuclear
matter, we analyze here whether the Atiyah-Manton scheme can be
used for introducing the proper variables to describe the skyrmion
system. As a multi-instanton solution, we adopt a modified 't
Hooft ansatz\cite{tHooft} to ensure that each instanton has a
finite size and relative orientations to others. This ansatz
provides us with the smallest set of dynamical variables for the
skyrmion system. We study the structure of the ground state as a
function of density by minimizing with respect to the parameters
describing the dynamical degrees of freedom. The method we used
can be considered as a variational method. Once we are able to
describe the ground state, we shall relax the parameters to go
beyond skyrmion matter.

This summary has served not only to present a brief report on the
status of skyrmion matter, but also to introduce our calculational
techniques. For completeness, we should mention the rational map
technique \cite{HMS98}, which has been very successful for baryons
up to $B = 27$. This technique has also been applied to skyrmion
matter, but only in two dimensions for which a Skyrme lattice with
hexagonal symmetry \cite{BS98} has been obtained. However,
applying this technique to skyrmion matter in three dimensions is
difficult in its present form, since, although the parameters of
this scheme may cover most of the configurations of the ground
state of finite baryon-number systems, their physical meaning is
obscure.

This paper is organized as follows; in the following section, we
describe how to apply the Atiyah-Manton ansatz to the skyrmion
crystals. The numerical results are presented in Section 3 and
Section 4 contains our conclusions.

\section{Ansatz for Skyrmion Matter}

\subsection{The Skyrme model Lagrangian}

Our starting point is the model of Skyrme \cite{skyrme} whose
Lagrangian density reads~\footnote{As a side remark, we should
clarify our point of view regarding the Skyrme Lagrangian
(\ref{skyrmeL}). For mathematicians, the Lagrangian
(\ref{skyrmeL}) $defines$ the problem and the goal for them is to
$solve$ the problem. This endeavor led to some beautiful results
for the structure of ``mathematical baryons" with the baryon
number $B$. For physicists who are interested in understanding the
physics of $B$-body systems, (\ref{skyrmeL}) is only an
approximation to nature and possibly a poor one for certain
processes. While the first term of (\ref{skyrmeL}) is the current
algebra term and hence is a legitimate ingredient of QCD, what
comes after that term is not generally known, and depends in
practice upon what one wants to study. It is understood that in
the limit of large number of colors, $N_C\rightarrow\infty$, QCD
can be represented in terms of the pion fields but in an infinite
series. For studying low-energy interactions with pions, one can
write the series systematically in terms of derivatives of the $U$
field; this leads to chiral perturbation theory. However if one
wants to study high energy processes or hadronic excitations under
extreme conditions (high temperature and/or high density), such a
large $N_c$ Lagrangian in the simplest form does not
work~\cite{HY-PR}. The same difficulty arises when one wants to
obtain the baryons as solitons from the effective Lagrangian.
There is no workable systematic way known to write down the
appropriate Lagrangian that can quantitatively describe baryons.
It is therefore more the reason for surprise that although there
is no theoretical justification from ``first principles", the
Skyrme Lagrangian (\ref{skyrmeL}) with the given quartic term
turns out to give qualitatively correct description of the
baryons~\cite{ZB86}. What we are doing here is that we simply
assume that (\ref{skyrmeL}) applies to nuclear matter as well and
take it (with the above caveat in mind) to study how to go from
the skyrmion matter structure to the nuclear matter structure that
we wish to describe. We are not concerned here with the accuracy
of the effective Lagrangian itself.  Our point of view is that
once we know how to do this with the Skyrme Lagrangian, we will be
able to do a realistic calculation once a Lagrangian that
represents QCD realistically is found.}
\begin{equation}
{\cal L} = \frac{f^2_\pi}{4} \mbox{Tr}(\partial_\mu U^\dagger
\partial^\mu U) + \frac{1}{32e^2} \mbox{Tr} ( \left[ U^\dagger
\partial_\mu U, U^\dagger \partial_\nu U \right]^2
),\label{skyrmeL}
\end{equation}
where $U(\vec{r},t)$ is the nonlinear realization of the pion
fields $\pi_i(\vec{r},t)$($i$=1,2,3); {\it viz.\/}
\begin{equation}
U(\vec{r},t) = \exp(i \vec{\tau} \cdot \vec{\pi}/f_\pi),
\end{equation}
with the Pauli matrices $\tau_i$($i$=1,2,3) generating the $SU(2)$
isospin space. Here, $f_\pi$ is the pion decay constant and $e$ is
the so called {\it skyrme parameter}. The pions are assumed
massless in this Lagrangian density.

It is more convenient and elegant to use dimensionless units for
the length and energy~\footnote{The following scaling is implied
by our choice of units: the lengths change as $[ef_\pi] x
\rightarrow x$ and the energies as  $E [e/(f_\pi 6\pi^2)]
\rightarrow E$} in terms of which the Lagrangian is expressed as
\begin{equation}
{\cal L} = \frac{1}{24\pi^2} \mbox{Tr}(L_\mu L^\mu)
+ \frac{1}{192\pi^2} \mbox{Tr} ( \left[ L_\mu,L_\nu \right]^2 ),
\end{equation}
where $L_\mu$ is the ``left current" defined as
\begin{equation}
L_\mu \equiv \partial_\mu U^\dagger U \equiv i
\vec{\tau}\cdot\vec{\ell}_i\ .
\end{equation}
The Faddeev-Bogomoln'y bound of a soliton solution carrying the
baryon number $B$ is~\footnote{Unlike in the case of the magnetic
monopole, the equality in (\ref{FBbound}) is never satisfied for a
nontrivial configuration for $U$.}
\begin{equation}
E \ge |B|,\label{FBbound}
\end{equation}
where $E$ is the energy of the static soliton solution
\begin{equation}
E = \frac{1}{12\pi^2} \int d^3 x  \left[ \vec{\ell}_i \cdot \vec{\ell}_j
+ \frac12 (\vec{\ell}_j \times \vec{\ell}_k )^2 \right].
\end{equation}
and the baryon number is given by
\begin{equation}
B =  \frac{1}{24\pi^2} \varepsilon^{ijk} \int d^3 x  \mbox{Tr}(L_i L_j L_k)
= \frac{1}{12\pi^2} \varepsilon^{ijk} \int d^3x
\vec{\ell}_i \cdot (\vec{\ell}_j \times \vec{\ell}_k).
\end{equation}

For the pion decay constant and the Skyrme parameter,
the following two sets of  values have been
widely used in the literature. One\cite{JR83} is
$$\left\{
\begin{array}{ll}
f_\pi = 93 \mbox{ MeV}, & \mbox{ the empirical pion decay constant,} \\
e = 4.75, & \mbox{ from a fitting to $g_A=1.25$.}
\end{array} \right.
$$
This choice leads to a value for the skyrmion mass $M_{B=1}=1.23
\times 6\pi^2 f_\pi/e \sim 1425$ MeV~\footnote{This looks much too
big for a nucleon. However as explained in \cite{NRZ}, there is a
term which is not included in \cite{JR83} of ${\cal O} (N_c^0)$
Casimir energy that comes out to be of order $\sim -500$ MeV.} and
the length unit becomes $197/(93 \times 4.75) \sim 0.45$ fm. The
other is obtained by a fit to the $N$-$\Delta$ mass
difference~\cite{ANW83} that arises after quantizing the classical
soliton to order $N_c$, to include $1/N_c$ effects, which requires
the following values

$$\left\{
\begin{array}{l}
f_\pi = 64.5,  \\
e = 5.45,
\end{array} \right.
$$
In this case, the length unit is $197/(64.5 \times 5.45)\sim 0.56$
fm.

\subsection{From instantons to skyrmions}

Atiyah and Manton\cite{AM89} proposed a remarkable ansatz, with
which a multi-skyrmion solution of the Skyrme Lagrangian carrying
the baryon number $N$ is expressed in terms of a multi-instanton
solution of charge $N$.\footnote{One should not confuse this
instanton solution with that of the color gauge field in QCD. We
are just borrowing the mathematical structure. Thus, differently
from the massive Yang-Mills theory, the presence of the pion mass
term in the Skyrme Lagrangian does not cause any problem in using
the 't Hooft instanton solution. } Explicitly, the soliton
solution is taken to be given by
\begin{equation}
U(\vec{x}) = C {\cal S} \left\{ {\cal P} \exp \left[
\int^\infty_{-\infty} - A_4(\vec x, t) dt \right] \right\}
C^\dagger,
\label{AM}
\end{equation}
where the time component of the gauge potential of an
instanton field of charge $N$ is integrated along the time
direction. Here, ${\cal P}$ denotes  the time-ordering,
${\cal S}$ is a constant matrix to make $U$ approach 1 at
infinity and $C$ describes an overall $SU(2)$ rotation.
The homotopy ensures that the static soliton configuration
carries the same baryon number as the total charge of the
instanton.

An exact solution for the multi-instanton of charge $N$ is known
and takes the form
\begin{equation}
A_4 (\vec{x},t) = \frac{i}{2} \frac{\vec{\nabla} \Phi}{\Phi}
\cdot \vec{\tau},
\label{A4}
\end{equation}
Here $\tau$'s are the Pauli matrices generating the {\em same}
$SU(2)$ space as those appearing in $U(\vec{r},t)$. The ansatz
reduces the equations of motion, $\partial_\mu F^{\mu\nu}=0$,
together with the self-duality relation,
 $\tilde{F}_{\mu\nu} = F_{\mu\nu}$, to
$\partial^2\Phi=0$. Two scalar functions $\Phi$, yielding
instanton charge $N$, are well-known in the
literature.

A first choice, known as 't Hooft instanton solution
\cite{tHooft}, is given by
\begin{equation}
\Phi = 1 + \sum^{N}_{n=1} \frac{\lambda^2_n}{(x- X_n)^2},
\label{tH}
\end{equation}
which contains $5N$ parameters for the positions $X_n^\mu
(\mu=1,2,3,4)$ of the $n$-th instanton in 4-dimensional Euclidean
space and their sizes $\lambda_n$. In the $N=1$ case, the
substitution of this 't Hooft instanton solution into the
Atiyah-Manton ansatz\cite{AM89} leads to the hedgehog skyrmion
solution
\begin{equation}
U(\vec{r}) = \exp(i \vec{\tau}\cdot\hat{r} F(r)),
\end{equation}
with the profile function given by
\begin{equation}
F(r) = \pi \left(1- 1/\sqrt{1+\frac{\lambda^2}{r^2} }\right).
\end{equation}
The minimum energy of the approximate solution comes out to be
$E=1.24$ when $\lambda=2.11$, which is very close to that of the
exact solution $E_{B=1}^{\mbox{\scriptsize exact}}=1.23$. When
substituted into the Atiyah-Manton ansatz, the 't Hooft instanton
solution provides $5N-1$\footnote{Note that in the time
integration procedure to obtain a skyrmion out of the ansatz one
can choose, for example, $X_{4,n=1}=0$ without loss of generality,
and therefore the skyrmion has one parameter less.} parameters to
the skyrmion system, these parameters becoming dynamical variables
in the procedure. This number is much smaller than the minimum
number of variables required, i.e., $6N$.

A second choice is the Jackiw-Nohl-Rebbi (JNR) instanton
solution\cite{JNR77}
\begin{equation}
\Phi = \sum^{N+1}_{n=1} \frac{\lambda^2_n}{(x- X_n)^2},
\label{JNR}
\end{equation}
which nontrivally generalizes the 't Hooft solution to contain 4
more parameters. This increase in the number of parameters
produces the toroidal minimum energy solution for the $B=2$
skyrmion\cite{LM94}. (Note that $5N+3 > 6N$ only when $N<3$.) On
the other hand, here, $X_n^\mu$ loses its meaning as the position
of the instanton so that it is quite difficult to guess the
geometry of the resulting skyrmion system. The number of
parameters is still less than $6N$.

\subsection{Instanton crystal and skyrmion crystal}
We proceed to modify the 't Hooft ansatz so as to be applicable to
many-skyrmion systems. In its present form, the $1/x^2$ tail of
the instanton, when summed over the infinite number of instantons
in a crystal, makes the $\Phi$ diverge and hence the 't Hooft
ansatz cannot be applied. To proceed, we shall slightly modify the
't Hooft multi-instanton structure, the time component of which,
$A_4(x)$, can be written explicitly as
\begin{equation}
A_4(\vec{x},t) = \frac12
\frac{ \displaystyle  \sum_{n=1}^N i\vec{\tau}\cdot
\vec{\nabla} \phi(x,X_n,\lambda_n) }
{\displaystyle 1 + \sum_{n=1}^N \phi(x,X_n,\lambda_n) },
\label{A40}\end{equation}
with
\begin{equation}
\phi(x,X_n,\lambda_n) = \frac{\lambda_n^2}{(x-X_n)^2}.
\label{phi0}\end{equation} The modification consists of
introducing the relative orientation of each instanton without
changing the instanton charge:
\begin{equation}
A_4(\vec{x},t) = \frac12 \frac{ \displaystyle  \sum_{n=1}^N C_n
i\vec{\tau}\cdot \vec{\nabla} \phi(x,X_n,\lambda_n) C^\dagger_n }
{\displaystyle 1 + \sum_{n=1}^N \phi(x,X_n,\lambda_n) },
\label{modifiedA4}\end{equation} with $C_n \in
SU(2)$($n=1,2,\cdots,N$). The $C_n$'s may be parameterized by the
Euler angles $(\alpha, \beta, \gamma)$,
\begin{equation}
C=e^{i\alpha \tau_z/2} e^{i\beta \tau_x/2}e^{i\gamma \tau_z/2}.
\end{equation}
We can further change the instanton profile function $\phi$ to be
finite ranged by multiplying it with a suitable ``truncation
function" $h(x,X_n)$; {\it viz.\/}
\begin{equation}
\phi(x,X_n,\lambda_n) = \frac{\lambda_n^2}{(x-X_n)^2} h(x,X_n).
\label{phi}\end{equation} We have checked numerically that,
as long as $h(x,X_n)$ satisfies the constraint, $h=1$, when
$x=X_n$,  and is finite at infinity, the baryon number of the corresponding
skyrmion field does not change. For simplicity we choose a
spatially truncated function such as
\begin{equation}
h(x,X_n) = \left\{
\begin{array}{ll}
\displaystyle \left(1 - \frac{|\vec{x}-\vec{X}_n|^2}{R^2}\right)^p, &
|\vec{x}-\vec{X}_n| < R, \\
0 & |\vec{x}-\vec{X}_n| > R,
\end{array}  \right.
\label{cutoff}\end{equation} which is continuous and
differentiable up to $p$-th order. In our numerical analysis,
we fix $p$ to 3.

Due to this modification, Eq.(\ref{modifiedA4}) is no longer an
exact solution of the instanton self-duality equation. However
since the Atiyah-Manton strategy amounts to borrowing the
topological structure of the instanton configuration via the
holonomy, we believe that the use of this {\em approximate}
solution would not lead to serious problems at the skyrmion level.
We note that this modified formula resembles that figuring in the
instanton liquid model\cite{CDG79} used to study QCD vacuum
structure\cite{IM81,DP84,Sh82}, for which the corresponding
$A_4(\vec{x},t)$ is of the form
\begin{equation}
A_4(\vec{x},t) = \frac12 \sum_{n=1}^N
\frac{ C_n i\vec{\tau}\cdot \vec{\nabla} \phi(x,X_n,\lambda_n) C^\dagger_n }
{1 + \phi(x,X_n,\lambda_n) }.
\label{ILMA4}\end{equation}
The difference is in the way the sums are performed.

Now, each instanton is described by 9 parameters:
$X^\mu_n$($\mu$=0,1,2,3) defines its position in 4-dimensional
Euclidean space, $C_n \in SU(2)$ the relative orientation,
$\lambda^2$ the its strength and $R$ the range. Except for the
temporal position of the instantons, the other instanton
parameters will preserve their physical meaning when transformed
to the skyrmion configuration with the particles well-separated.
Thus, it is rather easy to guess and construct an instanton
crystal corresponding to the skyrmion crystal of an expected
symmetry. Explicitly, the ansatz for a given instanton crystal
reads as
\begin{equation}
A_4(\vec{x},t) = \frac12
\frac{ \displaystyle \sum_{\mbox{\scriptsize box}}
\sum_{n=1}^N C_n i\vec{\tau}\cdot
\vec{\nabla} \phi(x,X_{n,\mbox{\scriptsize box}},\lambda_n) C^\dagger_n }
{\displaystyle 1 + \sum_{\mbox{\scriptsize box}}\sum_{n=1}^N
\phi(x,X_{n,\mbox{\scriptsize box}},\lambda_n) },
\end{equation}
where the summation runs now over instantons in a periodic box
and  {\em for all the boxes} as far as the instanton in that box
is within the range. The spatial periodicity of the crystal
requires that
\begin{itemize}
\item [i)]
all the image particles in each box share the same values of
the parameters for $\lambda_n^2$, $C_n$ and $X_{0,n}$;

\item [ii)]
only the spatial position parameters depend on the position of the box.
\begin{equation}
\vec{X}_{n,\mbox{\scriptsize box}} = \vec{X}_n + \mbox{ box position}.
\end{equation}

\end{itemize}

\begin{figure} 
\centerline{\epsfig{file=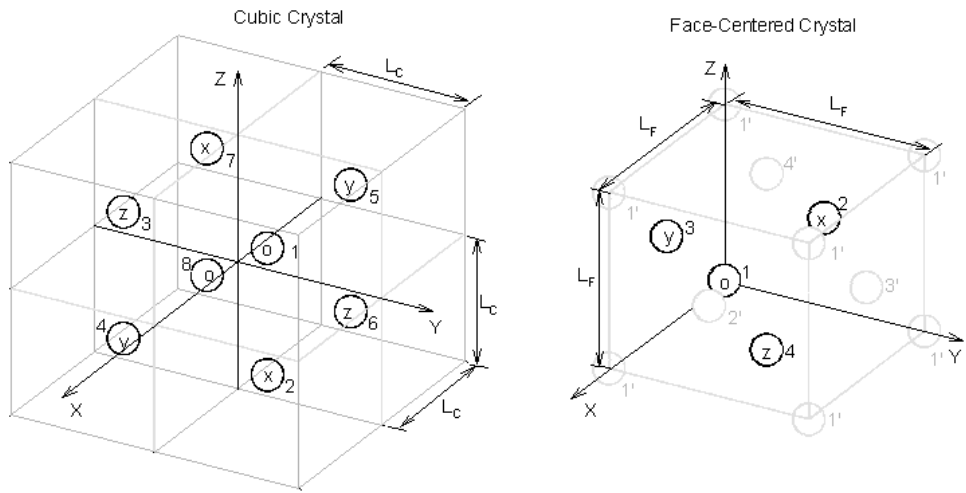,width=15cm,angle=0}}
\caption{Face centered instanton crystal. The location of each
instanton is denoted by a circle. The letters ``x", ``y" and ``z"
inside the circle mean that the instanton is rotated by $\pi$
around this axis. ``o" means unrotated instanton. The image
particles belonging to the neighboring boxes are presented by gray
circles and labelled as ``$i^\prime$", $i =1,2,...$. }
\end{figure}

We show in Fig.1 two of the crystals that can be constructed in
the way described. Figure 1a illustrates the simple cubic crystal
(CC) with the lattice constant $L_C$. This configuration carries
one unit of baryon number in a volume of $L_C^3$. The instanton
parameters that we have used in searching the ground state(s) are
listed in Table 1 and 2. The symmetry allows us to fix the size
and the range of each instanton to be equal. Although the ground
state is found for $T=0$ later, we impose the most  general
condition for the temporal parameter consistent with the CC
structure. Strictly speaking, it is not the conventional CC. As
far as the field $U(\vec{r})$ is concerned, exactly the same field
configuration is repeated with a period of $2L_C$. Thus, it is
more convenient to work with a box of length $2L_C$ containing
eight unit cells. However, with $T=0$ and for large $L_C$, each
unit cell shows the same baryon number distribution that forms a
CC. If $T\neq 0$, it is somewhat like an NaCl crystal.

\begin{table}
\begin{center}
\begin{tabular}{cccccccccc}
\hline
\# & $X_0$ & $X_1$ & $X_2$ & $X_3$ & $\alpha$ & $\beta$ & $\gamma$ & $\lambda$ & $R$ \\
\hline
1 & $-T/2$ & $+L_C/2$ & $+L_C/2$ & $+L_C/2$ & 0 & 0 & 0 & $\lambda$ & $R$ \\
2 & $+T/2$ & $+L_C/2$ & $+L_C/2$ & $-L_C/2$ & 0 & $\pi$ & 0 & $\lambda$ & $R$ \\
3 & $+T/2$ & $+L_C/2$ & $-L_C/2$ & $+L_C/2$ & $\pi$ & 0 & 0 & $\lambda$ & $R$ \\
4 & $-T/2$ & $+L_C/2$ & $-L_C/2$ & $-L_C/2$ & $\pi$ & $\pi$ & 0 & $\lambda$ & $R$ \\
5 & $+T/2$ & $-L_C/2$ & $+L_C/2$ & $+L_C/2$ & $\pi$ & $\pi$ & 0 & $\lambda$ & $R$ \\
6 & $-T/2$ & $-L_C/2$ & $+L_C/2$ & $-L_C/2$ & $\pi$ & 0 & 0 & $\lambda$ & $R$ \\
7 & $-T/2$ & $-L_C/2$ & $-L_C/2$ & $+L_C/2$ & 0 & $\pi$ & 0 & $\lambda$ & $R$ \\
8 & $+T/2$ & $-L_C/2$ & $-L_C/2$ & $-L_C/2$ & 0 & 0 & 0 & $\lambda$ & $R$ \\
\hline
\end{tabular}
\end{center}
\caption{Values of the input parameters for the CC structure ansatz}
\end{table}

Fig.1b shows a Face-Centered Crystal (FCC) with lattice constant
$L_F$. A box of volume $L_F^3$  contains 4 baryons. Here again,
the nearest skyrmions are rotated by $\pi$ with respect to the
axis perpendicular to the line joining the pair. The choice of the
other parameter is based on similar arguments to those of the CC
crystal. In the remaining part of this paper, we will only work
only with this FCC configuration. It fulfills the main purpose of
this paper, which is to show that our technique, described in the
previous section, is relevant to describe skyrmion matter.

\begin{table}
\begin{center}
\begin{tabular}{cccccccccc}
\hline
\# & $X_0$ & $X_1$ & $X_2$ & $X_3$ & $\alpha$ & $\beta$ & $\gamma$ & $\lambda$ & $R$ \\
\hline
1 & $-T/2$ & 0 & 0 & 0 & 0 & 0 & 0 & $\lambda$ & $R$ \\
2 & $+T/2$ & 0 & $+L_F/2$ & $+L_F/2$ & 0 & $\pi$ & 0 & $\lambda$ & $R$ \\
3 & $+T/2$ & $+L_F/2$ & 0 & $+L_F/2$ & $\pi$ & $\pi$ & 0 & $\lambda$ & $R$ \\
4 & $+T/2$ & $+L_F/2$ & $+L_F/2$ & 0 & $\pi$ &   0   & 0 & $\lambda$ & $R$ \\
\hline
\end{tabular}
\end{center}
\caption{Values of the input parameters for the FCC structure ansatz}
\end{table}

It is important to realize that we are putting the starting
crystal symmetry in our initial conditions in order to generate
the required crystal structure for the ground-state configuration.
In principle, all the parameters of our ansatz can be treated as
free parameters or variables without imposing any restriction.
This is the main advantage of our procedure over the more
conventional ones \cite{Kl85,JV88,CJJVJ89,KS89}. In the next
section, we will illustrate the power of the present approach by
evaluating the energy associated with the shift of a single
skyrmion from its stable position (i.e., ground state) keeping all
others fixed.

We are now ready to work out the Atiyah-Manton construction
Eq.(\ref{AM}) with the modified instanton configuration
(\ref{modifiedA4}). Time-ordering makes the construction somewhat
complicated. In order to simplify the calculation, we introduce a
skyrmion field generator $\tilde{U}(\vec{r})$ given by
\begin{equation}
\tilde{U}(\vec{r},t) \equiv {\cal S}
{\cal P} \exp \left(-\int^t_{-\infty} A_4(\vec{r},t^\prime)
dt^\prime \right),
\end{equation}
The generator satisfies the differential equation
\begin{equation}
\partial_t \tilde{U}(\vec{r},t) = - A_4(\vec{r},t)
\tilde{U}(\vec{r},t).
\end{equation}
The skyrmion ansatz $U(\vec{r})$ can be obtained
from the generator in the limit

\begin{equation}
\lim_{t\rightarrow \infty} \tilde{U}
\end{equation}

We proceed therefore by solving the differential equation,
which is a conventional time evolution problem, from
$t_0 = -\infty$ where it takes the initial value

\begin{equation}
\tilde{U}(\vec{r},-\infty)={\cal S},
\end{equation}
to the final time $t$. The constant matrix ${\cal S}$
is chosen so that
\begin{equation}
U(|\vec{r}|\rightarrow\infty) = 1.
\end{equation}

In the numerical procedure of evaluating physical quantities, we
only need the Sugawara variables $L_\mu$. They can be directly
obtained as a $t\rightarrow\infty$ limit of $\tilde{L}_\mu$
defined as
\begin{equation}
\tilde{L}_\mu =  \partial_\mu \tilde{U}^\dagger \tilde{U}
\end{equation}
which satisfies the differential equation
\begin{equation}
\partial_t \tilde{L}_\mu = [A_4(\vec{r},t), \tilde{L}_\mu]
- \partial_\mu A_4(\vec{r},t).
\end{equation}
The differential equation can be solved numerically by changing the
integration variable as $t=\tan\theta$, which makes the integration
range finite from $-\pi/2$ to $\pi/2$.

\section{Results}

Our calculation starts from an ansatz, at low density, constructed
from an FCC lattice with 4 skyrmions per site. At first, for a
given $L_F$, we search for the lowest energy configuration(s) by
adjusting the other parameters such as $\lambda$, $R$ and $T$.
Shown in Fig.2 is $E/B$ obtained in that way as a function of
$L_F$. As the density increases from the dilute skyrmion system,
the energy per baryon decreases to a minimum.

\begin{figure} 
\centerline{\epsfig{file=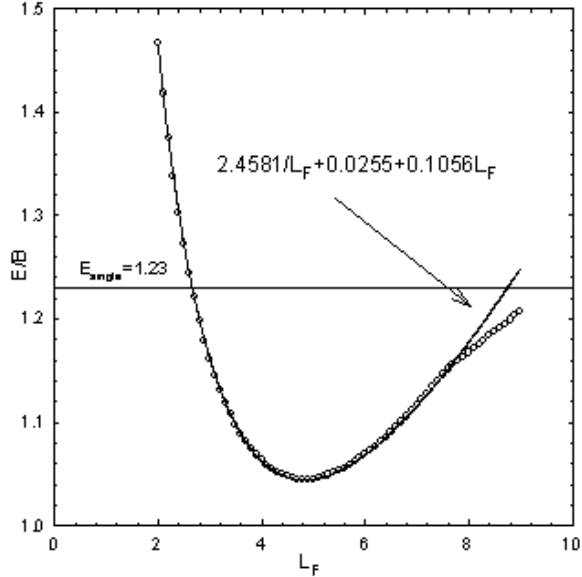,width=8cm,angle=0}}
\caption{$E/B$ as a function of $L_F$ the lattice constant of the
FCC crystal.}
\end{figure}

A best fit to this curve in the region around the minimum gives

\begin{equation}
E/B = 0.1056L_F + 0.0255 + 2.4581/L_F.
\end{equation}
Therefore
\begin{equation}
(E/B)_{min} = 1.044,
\end{equation}
and
\begin{equation}
L_{\min}= 4.82.
\end{equation}
This result is comparable to the lowest value $E/B=1.038$ obtained
by Kugler and Shtrikman\cite{KS89}.\footnote{For a rough
estimation, if we take $f_\pi=93$ MeV and $e=4.65$, our values
correspond to $\sim 2.18$fm and to $B/V=0.38$ nucleons/fm$^3$. On
the other hand, if we take $f_\pi=64.5$MeV and $e=5.45$, our
values correspond to $L_F\sim 2.72$fm and $B/V=0.20$
nucleons/fm$^3$. Thus, it is compatible to the normal nuclear
matter density.} This indicates that our calculation has reached
the true minimum of the skyrmion crystal at high density and that
the instanton parameters so determined can be taken as the
appropriate variables needed for describing the dynamics of the
multi-skyrmion system.

To see the configuration of the final skyrmion crystal, we present
in Fig. 3 the baryon density on the $x-y$ plane ($z=0$) at two
different values of $L_F$; (a) $L_F=8.0$ and (b) $L_F=5.0$. When
it is dilute (Fig.3a), the system is composed of well separated
single skyrmions lodged on each FCC lattice site. We may call this
the phase of ``single skyrmion FCC". At high density (including
the minimum energy configuration),  the individual skyrmion loses
its identity because of the overlap among the skyrmions. The dense
centers of the single skyrmion split into two and the final
configuration forms a CC crystal whose unit cells carry one-half
unit of baryon number. This can be interpreted as the
``half-skyrmion CC" obtained in Ref.\cite{KS89}. However, the
shape of the instanton tail described by Eq.(\ref{phi}) and the
cut-off function Eq.(\ref{cutoff}) are not accurate enough to
reproduce the exact half-skyrmion picture. Note that this
half-skyrmion phase has arisen without any additional changes in
our procedure: while our initial instanton configuration is still
FCC, the final skyrmion configuration turns out to be a CC lattice
with a fractional skyrmion per site.

\begin{figure} 
\centerline{\epsfig{file=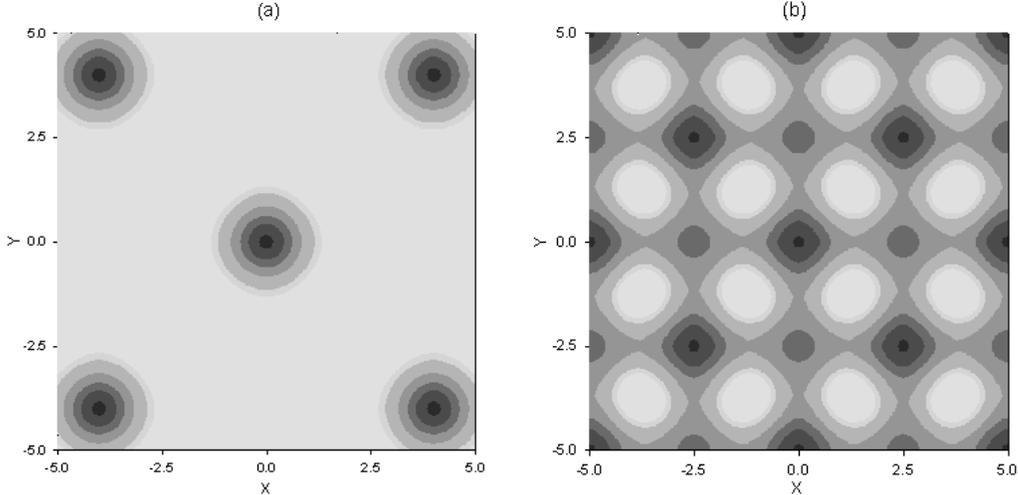,width=15cm,angle=0}}
\caption{The baryon number density of the FCC Skyrme crystal on
the $z=0$ plane at (a) $L_F=8.0$ and (b) $L_F=5.0$. The baryon
number density is normalized to the most dense region. Because of
the overlap, part of the baryon number  is accumulated at the
places in between the FCC lattice sites and  the configuration
becomes that of the half-skyrmion CC crystal.}
\end{figure}

A rough analysis on the configuration shows that the phase
transition from the single-skyrmion FCC to the half-skyrmion CC
takes place at around $L_F\sim 7.5$. Such a phase transition can
be explicitly seen in the parameter values that lead to the
minimum energy configurations at the given $L_F$ as shown in
Fig.4. For the high density phase ($L_F \le 7.5$), the parameters
vary around
$$\lambda/L_F\sim 1.1 \; , \; R/L_F \sim 1.4,$$ to
become a half-skyrmion CC, while for the low density phase ($L_F
> 7.5$) the parameters change sharply to
$$\lambda\sim 2  \; , \; R/L_F \sim 1.7$$
returning to the single skyrmion FCC crystal. The abrupt change
which one might interpret as a phase transition is also observed
in the $E/B$ curve of Fig.2, where the slope of the curve suddenly
changes to lower values around $L_F=7.5$ making the approach to
the asymptotic free value of $1.23$ slower.

\begin{figure} 
\centerline{\epsfig{file=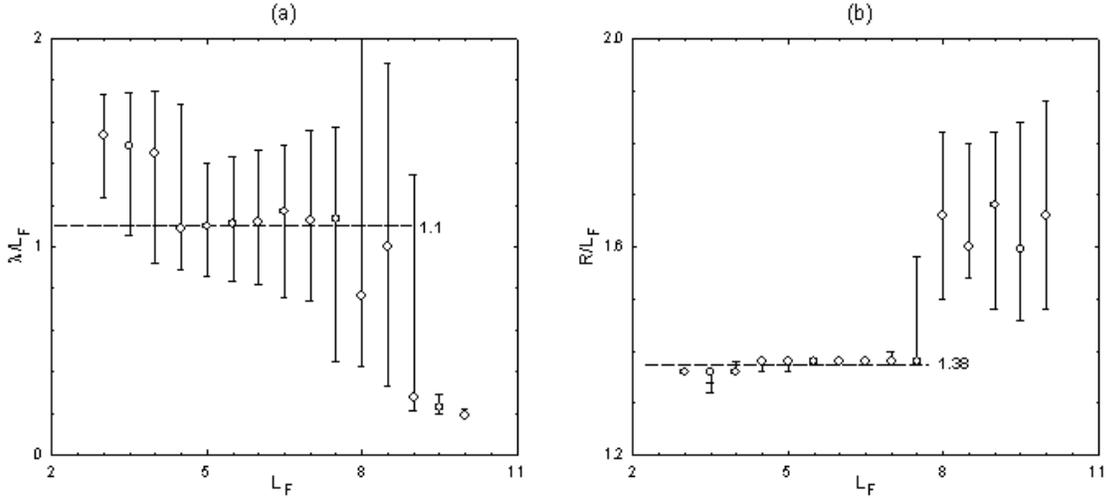,width=15cm,angle=0}}
\caption{The parameters (a) $\lambda/L_F$ and (b) $R/L_F$ of the
minimum energy configuration at the given $L_F$. The error bars
denote the parameter range that yields the energy within 5\% of
the corresponding minimum energy.}
\end{figure}

Though intriguing, it should however be pointed out that this
``phase transition" may not be physical. If we look at the
pressure defined as $P=\partial E/\partial V$, such a phase
transition happens where the pressure is negative. In the density
where the pressure is negative, the system is unstable and would
prefer a disordered phase to the symmetric FCC or CC
phases~\footnote{The phase transition that we have here should not
be confused with that of skyrmions in the $S_3$
sphere~\cite{mantonS,forkelS}. It happens in the range of density
where the pressure appears negative and this is caused because the
skyrmion system is constrained to be in FCC shape. By forming
disordered clusters of a few skyrmions, the system could develop
lower energy.}.

It is interesting to see that all the configurations in the
half-skyrmion phase show a certain scale invariance. In Fig.5 we
plot the baryon number density normalized to its maximum value
$\rho_B / \rho_{B,\mbox{\scriptsize max}}$ as a function of the
position normalized to the lattice constant $x/L$ along the
$y=z=0$ line. The figure shows that $\rho_B /
\rho_{B,\mbox{\scriptsize max}}$ for $L_F=7.5$ and $L=4.8$ depends
only on $x/L$. This type of scaling behavior repeats itself for
all values of  $L_F \le 7.5$, i.e., for high densities all the
configurations become almost half-skyrmion configurations. However
if $L_F \ge 7.5$ they lose this scaling property and become like
the initial FCC ansatz, i.e, a single skyrmion located at each FCC
site and with the overlapping region more or less void. This
scaling behavior may be related to the one found in the
variational approach of Kugler and Shtrikman\cite{KS89}: when the
configuration of the skyrmion crystal is expanded in harmonics,
the primary harmonics are found to play the dominant role. This
implies that at high density, the symmetry of the system dominates
over all dynamical details such as the interaction between two
nearest neighbors.

\begin{figure} 
\centerline{\epsfig{file=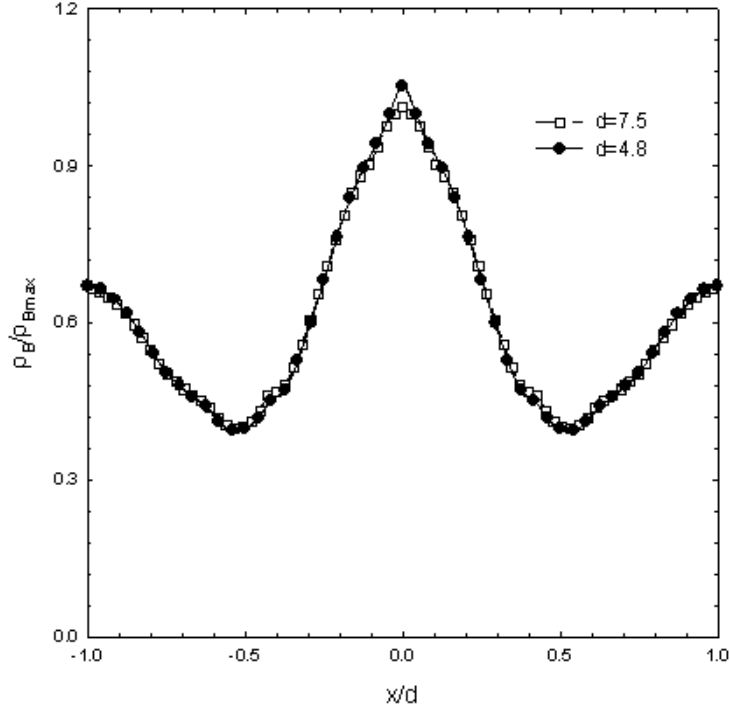,width=10cm,angle=0}}
\caption{$\rho_B / \rho_{B,\mbox{\scriptsize max}}$ as a function
of $x/L$ along the $y=z=0$ line.}
\end{figure}

As stressed, the main advantage of our approach is not in the
calculation of the ground state configuration but in the ability
to describe {\em with the same ansatz} the skyrmion system $far$
from the crystal structure without relying on symmetry arguments.
For example, we can calculate how the energy of the whole skyrmion
system changes when a single skyrmion is shifted from its stable
lattice position. This can be done just by shifting the position
of one instanton in Eq.(\ref{modifiedA4}) while leaving unchanged
that of all other instantons in the same box and in the
neighboring boxes (including its image particles). Since the
instantons are chosen to have a finite range, the calculation is
easy. Only the physical quantities inside this finite range are
affected. Shown in Fig.6 is the energy change of the system when a
skyrmion is shifted from its FCC lattice site by an amount $d$ in
the direction of the $z$-axis. The error bars are estimated by the
numerical fluctuations in the baryon number, assuming that the
energy is calculated with an error of the same magnitude. Two
extreme cases are shown. In the case of a dense system
($L_F=5.0$), the energy changes abruptly. For small $d$, it is
almost quadratic in $d$. It implies that the dense system is in
the crystal phase. On the other hand, in the case of a dilute
system ($L_F=10.0$), the system energy remains almost constant up
to some large $d$, which implies that the system is in a gas (or
liquid) phase. If we let all the variables vary freely, the system
will prefer to change to a disordered phase -- in which a few
skyrmions will form clusters -- to be more attractive.

\begin{figure} 
\centerline{\epsfig{file=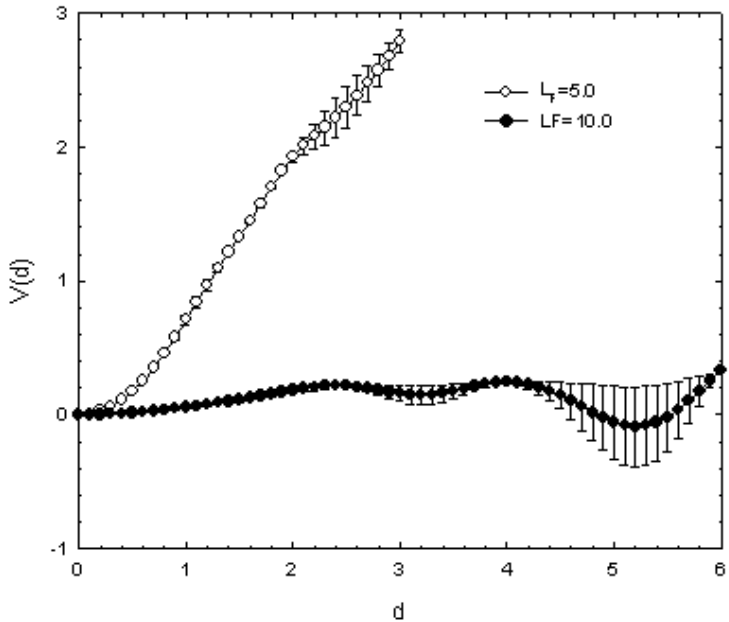,width=10cm,angle=0}}
\caption{The energy cost to shift a single skyrmion from its
stable position by an amount $d$ in the direction of the
$z$-axis.}
\end{figure}

\section{Conclusion}
In this paper, we propose a systematic way of introducing
dynamical variables in skyrmion matter that would allow us to
study nuclear matter with a link to large-$N_c$ QCD. The method
exploits the Atiyah-Manton ansatz to construct the pion field
configuration for the multi-skyrmion system from the
muti-instanton configuration. By modifying the 't Hooft instanton
solution, we can introduce a set of basic dynamical variables,
such as, the positions of the single skyrmions, the relative
orientations in the isospin space, sizes, etc. to cover the
necessary configuration space. To check whether these variables
can indeed cover sufficiently the important configuration space of
skyrmion matter, we look for the ground state of the system.
Although not perfect in its details -- e.g., the shape of the
tails of the instanton, the numerical results show that our
procedure is working satisfactorily.

The power of our approach is that we can go without much effort
beyond the ground state of the skyrmion crystal which is not
readily accessible to other approaches. As an example, we
demonstrate how easily one can evaluate the energy change of the
system when a single skyrmion is shifted from its stable position
while the others are kept fixed. A careful analysis of the data
enables us to investigate the vibrational energy spectrum of the
Skyrme crystal and hence its thermodynamic properties, e.g., the
specific heat, compressibility, etc. of skyrmion matter.

What is obtained in this paper sets the stage for the next step.
\begin{itemize}
\item
First of all, the theory can be quantized. Substituting in the
Atiyah-Manton skyrmion-instanton connection our modified 't Hooft
instanton ansatz, making the dynamical variables  ``{\em
time-dependent}," and plugging the result into the Skyrme
Lagrangian density, after integrating over the space variables, we
obtain the Lagrangian describing the skyrmion dynamics,
\begin{equation}
\int d^3 r {\cal L}(U(\vec{x},q_\xi(t))) = L(q_\xi,\dot{q}_\xi)
= \frac12 {\cal M}_{\xi\eta} \dot{q}_\xi \dot{q}_\eta - V(q_\xi),
\end{equation}
where $q_\xi(t)(\xi=1,2,\cdots,9N)$ denote the variables
introduced as $X_n$, $\lambda_n$ and ${\cal M}_{\xi\eta}={\cal
M}_{\xi\eta}(q_\zeta)$ are the inertia tensors conjugate to these
dynamical variables. This Lagrangian yields the equations of
motion for these variables. If these equations of motion are
solved by applying molecular dynamics simulation
techniques\cite{MD1}, we expect to see that the skyrmion crystal
melts to a liquid at a certain density.

\item By incorporating fluctuations in the pionic field\cite{fluc},
we should be able to study how pions (and also kaons) behave in dense
medium. This will provide a self-consistent scheme to study the
role of pions as Goldstone bosons in dense medium, in particular
in connection with Goldstone boson condensation and chiral
restoration. The presently available information on this matter is
incomplete in that although chiral symmetry is maintained, the
role of pions in the structure of baryons as described by e.g.
chiral Lagrangian is obscure particularly when the quark
condensate representing the QCD vacuum structure is modified by
density.
\end{itemize}
These two issues will be addressed in the forthcoming publication.

\section*{Acknowledgments}
Two of the authors (MR and VV) are grateful for the hospitality of
Korea Institute for Advanced Study where this work was performed.
V.V. thanks the University of Valencia for a travel grant. This
work was supported in part by SEUI-BFM2001-0262,
KOSEF-1999-2-111-005-5 and KRF-2001-015-DP0085.

\end{document}